\documentclass[prc,showpacs,twoside,twocolumn]{revtex4-1}
\usepackage{graphicx,amsmath,amsfonts}
\usepackage{amssymb,amscd,amsxtra,amsthm}
\usepackage{epsfig,bm,dcolumn}
\usepackage{epstopdf}
\usepackage{xcolor}

\makeatletter
\let\cat@comma@active\@empty
\makeatother

\begin{document}

\title{Nuclear transparency and shadowing of the two-step process in the $A(e,e'\pi^+)$ reaction}

\author{Tae Keun Choi}
\thanks{tkchoi@yonsei.ac.kr}
\affiliation{Department of Physics and Engineering Physics, Yonsei University, 
Wonju 26493, Korea}

\author{Kook-Jin Kong}
\thanks{kong@kau.ac.kr}
\affiliation{Research Institute of Basic Science, Korea Aerospace
University, Goyang 10540, Korea}

\author{Byung-Geel Yu}
\thanks{bgyu@kau.ac.kr}
\affiliation{Research Institute of Basic Science, Korea Aerospace University Goyang, 10540, Korea}



\begin{abstract}
We investigate nuclear transparency induced by pion electroproduction on nuclei, $A(e,e'\pi^+)$, with our interest in the role of the two-step process in the Glauber multiple scattering theory. Based on the framework in which the quantum diffusion model is incorporated into the Glauber theory, the experimental data in the range of photon virtuality $0.5< Q^2< 10$ GeV$^2/c^2$ are analyzed, including the recent JLab data at the 6 GeV electron beam together with the proposal for the 12 GeV upgrade.
Application of the quantum diffusion model during the formation length of the quark/antiquark $(q\bar{q})$ structure of a pion reproduces the increase in the nuclear transparency up to high $Q^2$, showing evidence of color transparency.
In contrast, through the coherent length of the $\gamma^*$ fluctuation into $\rho^0$, the two-step process with the $\rho N$ scattering cross section chosen to be $\sigma_{\rho N}=\sigma_{\pi N}$ contributes to a reduction of the nuclear transparency in the whole range of $Q^2$.
Combined with the one-step process (direct photon coupling) usually adopted in the original formulation of the Glauber theory, this reaction mechanism provides another source of the $Q^2$ dependence of the transparency at low virtualities through the shadowing in the initial state interaction.
By the further absorption due to the shadowing, a better agreement is obtained with the pion transparency measured in the  $A(e,e'\pi^+)$ reaction for $^{12}$C, $^{27}$Al, $^{63}$Cu, and $^{197}$Au nuclei.
A discussion of the $Q^2$ dependence of the parameter $\alpha$ for the transparency $T_A=A^{\alpha-1}$ is given.
\end{abstract}


\maketitle

\section{Introduction}

The study of nuclear transparency (NT) in nuclear reactions
induced by electromagnetic or hadronic probes serves as a testing
ground for understanding the medium effects in the reaction
mechanism at the QCD level.
NT originating from color transparency
(CT) of perturbative QCD quantifies the degree of the suppression
of the hadron interaction with the nuclear medium
\cite{Frankfurt:1992dx}.
According to QCD, which governs the
degrees of freedom of quarks and gluons, the $q\bar{q}$ structure
of produced hadrons at sufficiently high momentum transfers forms
a small-size configuration of color singlets whose size decreases
inversely with the momentum transfer $Q$ in the transverse
direction $r_\perp\approx 1/\sqrt{Q^2}$, while it undergoes the
Lorentz contraction in the longitudinal direction \cite{Frankfurt:1994hf}.
Thus, with less interaction with external color fields, just like the electric field of an electric dipole, it vanishes at distances much larger than the dipole size; the cross section of such a pointlike colorless object \cite{Brodsky1988} vanishes, $\sigma_{q\bar{q}}\propto r^2$ as $r\to 0$, leading to a suppression of the hadron interaction with the nuclear medium \cite{Blattel,kopel}.
The reduction in
color interaction is referred to as CT, which manifests itself as
an increasing $Q^2$ dependence of NT defined by
\begin{eqnarray}\label{nt0}
T_A={\sigma_A\over A\,\sigma_N}
\end{eqnarray}
for the nucleus of mass number $A$. It accounts for the normalized
ratio of cross sections for exclusive processes on nuclear and
nucleon targets \cite{Clasie,Qian}. On the other hand, in pion electroproduction the
colorless small-size configuration of the $q\bar{q}$ pair for the
produced pion could be a signature of the factorization of the
hard process of the $\gamma^*q\bar{q}$ coupling from the soft $\pi
NN$ vertex with generalized parton distributions by a suppression of the
interaction between the produced pion and the target nucleon
moving in opposite directions \cite{dutta0}. It is, therefore,
interesting to search for the onset of CT through hadronic
transparencies for the validity of QCD in various nuclear
reactions, such as those by a proton knockout and electromagnetic
productions of $\rho$, pion, and kaon
\cite{dutta0,bhet,kundu,makins,garrow,aira,cosyn1,fassi,Clasie,Qian,nuruzzaman}.

The pion transparency in the electronuclear reaction $A(e,e'\pi^+)$
has been measured at JLab using the 6 GeV electron beam
\cite{Clasie,Qian}. The transparency is described as a function of
$Q^2$ and the mass number $A$, showing clear evidence of CT by an
increasing $Q^2$ dependence in the range $Q^2 = 1.1\--4.7$
GeV$^2$. Conversely, the transparency at fixed $Q^2$ falls as $A$
increases. In this paper, we investigate pion transparency of the
reaction $A(e,e'\,\pi^+)$ based on the Glauber multiple scattering
theory \cite{Glauber}. Since the initial and final states of the
nucleus are not the same due to the charged pion production, only
the incoherent nuclear reaction can contribute to the NT. The
conventional approach to the Glauber theory relies solely on the
density function of nucleons and the scattering cross section of
pions off nucleons in nuclei. Thus, the pion transparency remains
constant regardless of increasing $Q^2$, since there is no $Q^2$ dependence
either in the Glauber theory or in the $\pi N$ cross section. Furthermore, the
predicted values are smaller than what is observed in experiments
\cite{Clasie,Qian}. This discrepancy indicates that additional
mechanisms should be explored beyond the traditional approach.

Larson {\it et al.} \cite{Larson} included a semiclassical
formula for the final state interaction (FSI), while Cosyn {\it
et al.} {\cite{Cosyn} considered a relativistic version of the
Glauber multiple scattering theory. Both studies incorporated the
effect of CT using the quantum diffusion model (QDM)
\cite{Farrar:1988me} with the empirical value of $\pi N$
scattering cross section $\sigma_{\pi N}$ from the Particle Data
Group (PDG). Kaskulov {\it et al.} \cite{Kaskulov} focused
on the production mechanism of the subnuclear reaction
$p(e,e'\pi^+)n$ to apply to the coupled-channels treatment for the
pion-nucleus interaction. By separating the longitudinal and
transverse cross sections on the nucleon when calculating the
elementary pion production cross section, the CT from the
transverse and longitudinal contributions was discussed. 
Larionov {\it et al.} \cite{Larinov} estimated the $\pi$ CT in the
reaction $\pi^- n\rightarrow\ell^{+}\ell^{-}n$ to provide
information complementary to the $\gamma^* p\rightarrow\pi^+ n$
process. Das \cite{Das} investigated the pion transparency,
incorporating the CT of the produced pions and the short-range
correlation (SRC) between nucleons in the nucleus into the Glauber
theory. The comparisons between the results with and without the
SRC have been given for each pion transparency in the range of
$Q^2 = 1.1\--9.5$ GeV$^2$ \cite{Cosyn,Kaskulov}.

In hadron-nuclear reactions exemplified above, an apparent increase in
transparency with increasing $Q^2$ is obtained by QDM
\cite{Farrar:1988me}. In this QCD-inspired model, it is deduced
that the initial small-size configuration of a pion after
electroproduction still proceeds to a physical meson as a
$q\bar{q}$ pair. Therefore, the strength of $\pi N$ scattering in
the nucleus decreases during the passage of the $q\bar{q}$ state,
called the formation length $l_F$, giving an increase in nuclear
transparency.
Nevertheless, the application of QDM leads to an overestimation of JLab data of Refs. \cite{Clasie,Qian} for heavier nuclei, which should be improved.
In this paper, for consistency with
experimental results, we introduce a new reaction mechanism that
plays the role of nuclear shadowing in the Glauber description of
electroproduction of mesons rather than refitting the parameters of QDM to the data.
Although the short range correlation (SRC) is favored in Refs. \cite{kundu,Das,Lee},
we focus on the role of the new entry in the present reaction $A(e,e'\pi^+)$ without SRC, since it provides only a constant contribution with its parameters too sensitive to demonstrate the significance of the absorption by shadowing.

In addition to the direct photon coupling to the $\pi
N$ interaction in electroproduction, Yennie \cite{Yennie70}
and Bauer \cite{Bauer78} considered the effect of virtual
photons fluctuating into vector mesons, which is called the
two-step process in the initial state interaction (ISI). It gives an
improvement to the Glauber theory because the $q\bar{q}$
fluctuation of a virtual photon into a vector meson affects the
reduction of the nuclear transparency along the coherent path
$l_c$, as will be discussed below.

This paper has been organized as follows.
In Sec. II, the Glauber scattering theory for analyzing the dependence of nuclear transparency in the reaction $A(e,e'\pi^+)$ on the virtuality $Q^2$ and the mass number $A$ is presented.
It is also demonstrated how to treat shadowing by the two-step process in the Glauber theory in the description of the nuclear transparency.
In Sec. III the QDM is included in the Glauber theory with the two-step process employed in the previous section. The calculations are compared to available experimental data at the JLab 6 GeV electron beam. Predictions are made for nuclear transparency in future experiments like the 12 GeV upgrade electron beams, and discussion of the theoretical improvement is presented.
Finally, in Sec. IV the key results are summarized.

\section{Formalism}

In the Glauber theory for the nuclear cross section of electropion
off a nucleus of mass number $A$, the underlying reaction is pion
electroproduction on a nucleon \cite{cky15},
\begin{equation}
\gamma^*(k) + N(p) \rightarrow \pi^+(p_\pi)+ N'(p'),
\label{eq:elecprod}
\end{equation}
which is described by the dependence of the Lorentz invariants,
$Q^2$, $W$, and $t$ with the photon virtuality $k^2=-Q^2$, the
total energy $W=\sqrt{M_p^2+2M_p\nu-Q^2}$, and $\nu=E_e-E'_e$ is
the energy of the virtual photon by the difference between the
initial and final electron beam energies. $M_p$ is the proton
mass, and $x_B=Q^2/2M_p\nu$ is the Bjorken scaling variable.
%
In Table \ref{tb1}, we list the
kinematical variables relevant to pion electroproduction in Eq.
(\ref{eq:elecprod}) in the range of photon virtuality $0.5<Q^2<10$ GeV$^2/c^2$, which covers the recent JLab data at 6 GeV electron beams \cite{Clasie,Qian} as well as the extension to 12 GeV proposed in Ref. \cite{dutta107}.

\begin{figure}[t]
\centering
\includegraphics[width=8.5cm]{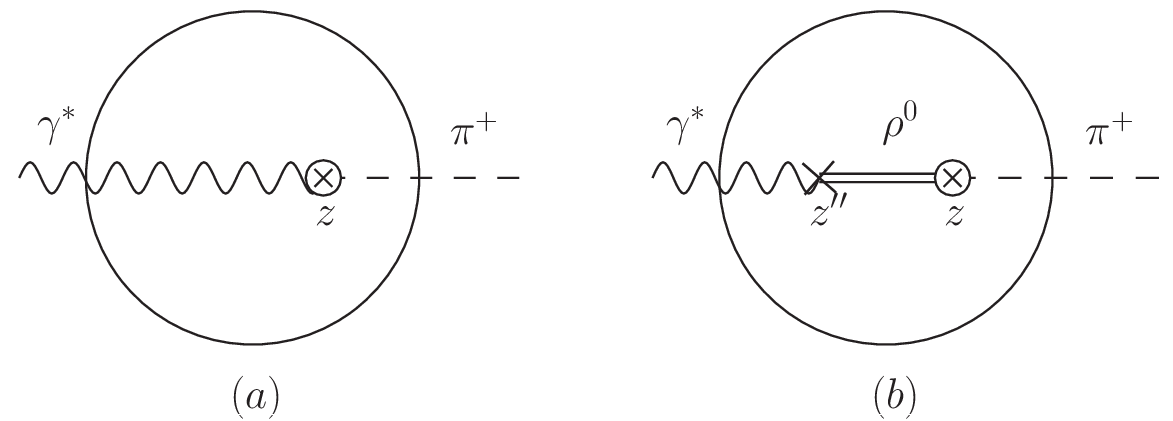}
\caption{ One-step and two-step processes to forward incoherent
$\pi^+$ electroproduction on the nucleus. The symbol $\otimes$
represents the interaction of $\gamma^*$ with $\pi N$ at the
position $z$ and the symbol $\times$ the $\gamma^*$-$\rho^0$
conjunction at the $z''$. (a) One-step $\pi^+$ production. (b)
Two-step $\pi^+$ production intermediated by $\rho^0$ meson. }
\label{fig1}
\end{figure}

We begin with the ratio $A_{eff}$ of nuclear to nucleon cross sections defined as \cite{Yennie70},
\begin{eqnarray}\label{aeff}
{d\sigma^{inc}_A\over dt}(\gamma^*A\to \pi^+A')=A_{eff}{d\sigma_{N}\over dt}(\gamma^*N\to\pi^+N'),
\end{eqnarray}
where $d\sigma^{inc}_A/dt$ is the incoherent cross section of the
reaction for the target nucleus $A$ to transit to the final
nucleus $A'$, and $d\sigma_N/dt$ is the hadronic cross section of
the electropion production off a nucleon in free space with
\begin{eqnarray}{}
&&A_{eff}=\int\limits^\infty_0 d^2b
\!\int\limits^\infty_{-\infty}\ dz \,\varrho(\bm{b},z)
\exp\Bigl({-} \sigma_{\pi N}\int\limits^\infty_z 
dz'\,\varrho(\bm{b},z') \Bigr)\nonumber\\
&&\hspace{1cm}\times \Biggl|1-\int^z_{-\infty}dz''\varrho(\bm{b},z'')\frac{\sigma_{\rho N}}{2}(1-i\alpha_{\rho N})e^{iq_L(z''-z)} \nonumber\\
&&\hspace{1cm}\times \exp\Bigl[-\frac{\sigma_{\rho N}}{2}(1-i\alpha_{\rho N})\int^z_{z''}dz'''\varrho(\bm{b},z''')\Bigr]\Biggr|^2.
\label{eq:nt1}
\end{eqnarray}
Then, the NT given in Eq. (\ref{nt0}) is written as \cite{Bauer78,Larinov}
\begin{equation}
T_A={1\over A}A_{eff}
\label{eq:nt}
\end{equation}
with the integration in the $A_{eff}$ performed over the impact
parameter $\bm{b}$ and the coordinate $z$ along the direction of
the incident photon. The nuclear density function is denoted by
$\varrho(\bm{b},z)$, which is normalized to the total number of
nucleons $A$. $\sigma_{\pi N}$ is the total cross section for $\pi^+ N$ scattering
soon after the pion production point $z$,
and $\sigma_{\rho N}$ is the total cross section proceeding up to
the position $z$ as denoted in Fig. \ref{fig1}(b).
$\alpha_{\rho N}$ is the ratio of the real part to the imaginary part
of the $\rho^0 N$ scattering amplitude at forward angles
\cite{sibirtsev}. The exponential term with the factor
$\sigma_{\pi N}$ in Eq.~(\ref{eq:nt1}) describes the attenuation of
pions passing through the nuclear thickness. Those terms in the
square of the absolute value signify the presence of two
interfering waves. The first term with the factor 1 corresponds to
the virtual photon incident on the nucleon at the position
($\bm{b},z$) for the direct photon
coupling in the original Glauber theory, as shown in Fig. \ref{fig1}(a).
In contrast, the integral
term, we call the two-step process as shown in Fig. \ref{fig1}(b), corresponds to the case of the fluctuation of the virtual
photon into $\rho^0$ meson incoming there \cite{Yennie70,Bauer78}.

In the two-step process, while the transverse component is
negligible in the eikonal scattering, the longitudinal component
of the momentum transfer to the nucleon in Eq. (\ref{eq:nt1}) is
given by the difference in the longitudinal components of the
momentum between the incident photon and the $\rho$ meson
\cite{Kaskulov},
\begin{eqnarray}\label{lc}
q_L=\sqrt{\nu^2+Q^2}-\sqrt{\nu^2-m_\rho^2}\approx{(Q^2+m_\rho^2)/ 2\nu}.
\end{eqnarray}
(see also Ref. \cite{Yennie70}, Eq. (53) in pp. $341\--342\,$). To
access the effect in association with the coherence length, which
is defined to be $l_c = 1/q_L$, we calculate the incoherent cross
section in Eq. (\ref{aeff}) at high energy, where we let the limit
$q_L \to 0$. In this limit, the phase factor
exp$[\,iq_L(z''-z)\,]$ can be dropped in Eq. (\ref{eq:nt1})
\cite{Bauer78}, and the second term for the two-step process is
further simplified as $ 1-{\rm exp}\Bigl[-\frac{1}{2}\sigma_{\rho
N}(1-i\alpha_{\rho N})\int\limits^z_{-\infty} \varrho(\bm{b},y)\,
dy\Bigr] $, and the full expression for $T_A$ is now given by
\begin{eqnarray}
&&T_A= {1\over A}\int\limits^\infty_0 d^2b
\!\int\limits^\infty_{-\infty}\ dz \,\varrho(\bm{b},z)\nonumber\\
&&\hspace{0.5cm} \times \exp\Bigl[{-} \sigma_{\pi
N}\int\limits^\infty_z dy\,\varrho(\bm{b},y)\, -\sigma_{\rho
N}\int\limits^z_{-\infty} dy\,\varrho(\bm{b},y)\, \Bigr].
\label{eq:nt2}
\end{eqnarray}

Note that the final result for the NT for the reaction
$A(\gamma^*,\pi^+)$ with the two-step process considered no longer
depends on the ratio $\alpha_{\rho N}$. The first term implements the attenuation of the
produced pion during the FSI, which begins at the lower bound $z$
of the pion production point in the first integral and extends
through the formation length $l_h$ [see Eq. (\ref{lf}) below]. At the QCD level, as
illustrated in Ref. \cite{howell}, the second term in Eq.
(\ref{eq:nt2}) represents the effect of the virtual photon
$\gamma^*$ fluctuating into a $\rho^0$ meson while traveling a
coherent length $l_c\,$. In
this context, the two-step term accounts for the shadowing effect
due to $\rho^0$ propagation within the ISI along the coherent
length $l_c$, up to the upper bound $z$ in the second integral. As
a result, the lower bound of the second integral, originally set
at $-\infty$, is replaced by the effective point $z - l_c\,$. This
replacement allows the initial shadowing process to have a
dependence on $Q^2$, as described in Eq. (\ref{lc}).

For the nuclear density $\varrho(r)$,
the Wood-Saxon distribution function is employed for all nuclei considered here \cite{sibirtsev},
\begin{equation}
\varrho (r) =\frac{\varrho_0}{1+\exp{[\, (r-R)/d\, ]}}\,,
\label{eq:ws}
\end{equation}
where the nuclear radius is parametrized as
$R=1.28A^{1/3}-0.76+0.8A^{-1/3}$ fm, and the diffusion parameter
is given by $d{=}1/\sqrt{3}$~fm.

\section{Data analysis}

\begin{table}[t]
\caption{ Kinematical variables $Q^2$ (GeV$^2/c^2$), $W$ (GeV),
$E_{e}$ (GeV), $\theta_e$ (deg), $\nu$ (GeV), $x_B$, and
$|\bm{p}_\pi|$ [GeV/c] in the pion electroproduction experiment at
JLab \cite{Clasie,Qian} and quantities relevant to pion
transparency in the range of $0.5\leq Q^2\leq 10$ GeV$^2/c^2$. The
coherent length $l_c$ (fm) in Eq. (\ref{lc}) and the formation
length $l_h$ (fm) in Eq. (\ref{lf}) are displayed. The total cross
section $\sigma_{\pi N}$ (mb) in the last column is chosen as the
average value read from the PDG database for $\pi^+p$ scattering
in the given range of pion momentum $p_\pi$ (GeV/c). In the
calculation of the two-step process we use the relation
$\sigma_{\rho N}=\sigma_{\pi N}$ from Ref. \cite{drell1966}.}
\begin{ruledtabular}
\begin{tabular}{llllllllll}
$Q^2$& $W$  & $E_{e}$&$\theta_e$& $\nu$&$x_B$&$|\bm{p}_\pi|$& $l_c$ &$l_h$&$\sigma_{\pi N}$ \\
\hline
0.55 &2.28  &3.4  &26   &2.6 &0.11&2.59 &0.9 &1.46& \\
0.69 &2.25  &3.5  &27   &2.6 &0.14&2.58 &0.8 &1.46& \\
0.77 &2.25  &3.68 &26   &2.64&0.16&2.62 &0.76&1.48& 30\\
0.92 &2.26  &3.85 &27   &2.75&0.18&2.73 &0.71&1.54& \\
1.05 &2.26  &4.02 &27   &2.82&0.2 &2.79 &0.67&1.57& \\
\hline
1.10 &2.26  &4.02 &27.76&2.83&0.21&2.8  &0.66&1.58& \\
2.15 &2.21  &5.01 &28.85&3.28&0.35&3.2  &0.47&1.80& \\
3.0  &2.14  &5.01 &37.77&3.58&0.45&3.43 &0.39&1.93& 28\\
3.91 &2.26  &5.77 &40.38&4.34&0.48&4.16 &0.38&2.34& \\
4.69 &2.25  &5.77 &52.67&4.73&0.53&4.49 &0.35&2.53& \\
\hline
5.0  &2.43  &11.0 &16.28&5.33&0.5 &5.12 &0.38&2.88& \\
6.5  &2.74  &11.0 &22.13&6.99&0.5 &6.78 &0.39&3.82& 25\\
8.0  &3.02  &11.0 &32.37&8.66&0.49&8.45 &0.40&4.76& \\
9.5  &3.09  &11.0 &47.71&9.68&0.52&9.43 &0.38&5.31& \\
\end{tabular}
\end{ruledtabular}
\label{tb1}
\end{table}

In this paper, we investigate the $Q^2$ and $A$ dependence of $T_A$ in an extended range of $Q^2$ wider than the existing JLab data at 6 GeV electron beams to provide information for the experiment at the 12 GeV upgrade \cite{dutta107}.
In particular, to avoid nucleon resonances contributing to the total cross section $\sigma_{\pi N}$, we select photon virtuality $0.5<Q^2<1.1$ GeV$^2/c^2$ in Table \ref{tb1} with the corresponding energy $W$ above the nucleon resonance region $W>2$ GeV and the variable $x_B$ chosen to be of the same order $10^{-1}$ as others for consistency.

The experimental data at JLab show that $T_A$ increases with increasing $Q^2$, indicating color transparency. However, it is known that the original Glauber theory with $\sigma_{\rho N}=0$ in Eq. (\ref{eq:nt2}) and the $\sigma_{\pi N}$ cross section read from Table \ref{tb1} yield a result inconsistent with the observation in experiments, staying nearly constant against the increase in $Q^2$, and even underestimating the data \cite{Clasie,Qian}.

Since the $q\bar{q}$ structure moves to the final $\pi^+N$ state
until it forms the physical $\pi^+$, such a QCD effect on the
reaction, which we call the QDM, should account for the increasing slope of $T_A$ in the
$Q^2$ dependence of the experimental data.
Following the construction of the QDM in Refs.
\cite{Farrar:1988me,Larson,Frankfurt:1994hf}, it is included in
the Glauber theory by replacing the $\sigma_{\pi N}$ in free space
with the effective cross section $\sigma_{\rm eff}$,
\begin{eqnarray}
&&\sigma_{\rm eff}(z',p_\pi)= \sigma_{\pi N}(p_\pi)
\Bigg[\theta(z'-l_h)
\nonumber\\&&\hspace{1.8cm}
+\bigg\{\frac{n^2\langle  k_t^2
\rangle}{Q^2}\left(1-\frac{z'}{l_h}\right) +\frac{z'}{l_h}\,
\bigg\}\theta(l_h-z') \Bigg], \ \ \ \ \
\label{effc}
\end{eqnarray}
which is now a function of the path length $z'=z-y$ of the
distance the $q\bar{q}$ pair travels after electroproduction.
Thus, it is to be integrated in Eq. (\ref{eq:nt2}) when QDM is considered.
The values of the cross section $\sigma_{\pi N}$ are read from the PDG database
for the total cross section for $\pi^+p$ scattering corresponding to pion
momentum $|\bm{p}_\pi|$, as presented in Table \ref{tb1}. Here,
$n=2$ is the number of valence quarks and antiquarks for the pion.
$\langle k_t^2 \rangle^{1/2}\simeq 0.35$ GeV$/c$ is the transverse momentum of a quark in a pion.
The formation length $l_c$ is the distance through which the $q\bar{q}$ pair travels soon
after the interaction until it forms the physical pion in
the final state. It is estimated by the uncertainty in time,
$\Delta t=(\sqrt{\bm{p}_\pi^2+m_{\pi'}^2}-\sqrt{\bm{p}_\pi^2+m_{\pi}^2})^{-1}$,
between the mesons having the same
quantum numbers superimposed with the produced pion \cite{dutta0,howell},
\begin{eqnarray}\label{lf}
l_h\approx {2|\bm{p}_\pi|\over \Delta M^2} 
\end{eqnarray}
with $\Delta M^2=m_{\pi'}^2-m_{\pi}^2$.
We adopt $|{\bm p}_{\pi}|$ in Table \ref{tb1} and $\Delta
M^2=0.7$ GeV$^2$, which is widely applied in the relevant
reactions \cite{Kaskulov,Das}. In Table \ref{tb1} we present the
specific lengths $l_c$ and $l_h$ featuring the initial shadowing
by the $q\bar{q}$ fluctuation of $\gamma^*$ into $\rho^0$ and the
formation of a physical pion in the final state.

\begin{figure*}[]
\centering
\includegraphics[width=13cm]{fig2.eps}
\caption{
Nuclear transparency $T_A$ vs $Q^2$ in the reaction $A(e,e'\pi^+)$ on four different nuclei: ${}^{12}$C, ${}^{27}$Al, ${}^{63}$Cu, and ${}^{197}$Au.
A discontinuity appears due to the use of three different sets of kinematic variables in the range $0.5< Q^2<10$ GeV$^2/c^2$ \cite{dutta0,dutta107}, as listed in Table \ref{tb1}.
$T_A$ in the low $Q^2$ region $0.55\leq Q^2\leq 1.05$ GeV$^2/c^2$ is investigated for testing the role of the shadowing by the two-step process.
$T_A$ in the range $1.1\leq Q^2\leq4.69$ GeV$^2/c^2$ reproduces the JLab data at 6 GeV electron beams with the prediction for the 12 GeV upgrade electron beam up to $Q^2=9.5$ GeV$^2/c^2$.
The dotted curve is from the original Glauber theory (GT), and the dashed curve is from GT + QDM.
The dash-dotted curve represents the GT+two-step process showing the effect of the shadowing in Eq. (\ref{eq:nt2}), which exhibits $Q^2$ dependence at low $Q^2$.
The cross section $\sigma_{\rho N}=\sigma_{\pi N}$ is chosen for each $Q^2$ range \cite{drell1966}.
The solid curve corresponds to our result from the shadowing in the ISI incorporated with the GT+QDM.
Data are taken from Refs. \cite{Clasie,Qian}. }\label{fig2}
\end{figure*}
\begin{figure*}[]
\centering
\includegraphics[width=13cm]{fig3.eps}
\vspace*{-2mm}
\caption{
Nuclear transparency $T_A$ vs mass number $A$ at fixed $Q^2$ in the reaction $A(e,e'\pi^+)$.
The solid and dash-dash-dotted curves correspond to the predictions with and without the shadowing in the GT+QDM. The respective contributions of QDM, and shadowing by the two-step process, are shown at $Q^2=1.1$ GeV$^2/c^2$ with the notation for the curve the same as in Fig. \ref{fig2}.
The $T_A$ in the range $1.1\leq Q^2\leq 4.7$ GeV$^2/c^2$ reproduces the JLab data for the 6 GeV electron beam.
$T_A$ in the lower and higher virtualities, varying from $Q^2=0.55$ to 9.5 GeV$^2/c^2$, are presented for testing $T_A$ in the wider range of $Q^2$.
Data are taken from Ref. \cite{Qian}.
}\label{fig3}
\end{figure*}
\begin{figure}[]
\centering
\vspace*{1mm}
\includegraphics[width=8.5cm]{fig4.eps}
\caption{
$Q^2$ dependence of the parameter $\alpha$ from the global nuclear data set of the $A(e,e'\pi^+)$ experiment at JLab.
Notations for the curves are the same as in Fig. \ref{fig2}.
The thick shaded area around $\alpha\approx0.76$ is extracted from pion-nucleus scattering \cite{carroll79}.
Data in the range $1.1\leq Q^2\leq 4.7$ GeV$^2/c^2$ are taken from Ref. \cite{Qian}.
}\label{fig4}
\end{figure}

All calculations are carried out in the laboratory system where
the nuclear target is assumed to be stationary and the nucleons
inside are also assumed to be stationary. In reality, the nucleons
inside the nucleus will be in Fermi motion, but their energy can
be negligible compared to that of the nucleus, and hence we
neglect its effect. This is equivalent to the application of the
proton-on-shell model in the quasifree approximation of the
struck proton, as discussed in Refs. \cite{Clasie,Qian}.
Since the scattering cross sections from protons and neutrons are nearly
identical, it is assumed that all targets within the nucleus are
protons, with the distinction between protons and neutrons being
ignored.

\subsection*{$Q^2$ and $A$ dependence of $T_A$}

Given the kinematical variables of $p(\gamma^*,\pi^{+})n$ in
Table \ref{tb1}, the behavior of the shadowing and QDM is tested
in the Glauber theory for NT at high photon virtualities,
including the JLab 12 GeV upgrade electron beams in Fig.
\ref{fig2}.
$T_A$ is presented as a function of $Q^2$ for four target nuclei, $^{12}$C, $^{27}$Al, $^{63}$Cu, and $^{197}$Au.
In each panel, $T_A$ is shown up to $Q^2=10$ GeV$^2/c^2$ in the range divided by three intervals of $Q^2$, as specified in Table \ref{tb1} with the JLab data of Ref. \cite{Qian} for comparison with the theoretical prediction.
As mentioned, without QDM and shadowing by the two-step process, the dotted curve resulting from the original GT fails to reproduce the $Q^2$ dependence of $T_A$.
The recovery of a linear dependence of $Q^2$ is achieved in the $T_A$ by using QDM, which leads to the dashed curve when combined with the original GT prediction with the cross section $\sigma_{\pi N}$ from Table \ref{tb1}.
While the agreement with the slope of the data should be a crucial indicator of CT, the result of GT+QDM instead results in an overestimation of the theoretical prediction.
To agree with the experimental data on $T_A$ by reducing the theoretical prediction so far, we now turn on the effect of the two-step process in Eq. (\ref{eq:nt2}) as illustrated in Fig. \ref{fig1}(b), with an expectation of the shadowing from the last integral term.
As a parameter, we assume that $\sigma_{\rho N}=\sigma_{\pi N}$ throughout the calculation by following the analysis of Ref. \cite{drell1966}.
The choice of $\sigma_{\rho N}$ found in Ref. \cite{cosyn1} further supports our assumption.
Given the $Q^2$ dependence from the lower limit of the integral, $z- l_c$, with $l_c=2\nu/(Q^2+m_\rho^2)$ in Eq. (\ref{eq:nt2}), the dash-dotted curve represents the shadowing by the two-step process in the GT with $\sigma_{\pi N}$ taken from Table \ref{tb1} without QDM.
While diminishing the contribution of GT+QDM, our result is given by the solid curve from GT+QDM+two-step process in Fig. \ref{fig2}.

Thus, within the present formalism for NT, the $Q^2$ dependence arises not only from the absorption by QDM in the FSI but also from the shadowing by the two-step process (via the $q\bar{q}$ fluctuation of $\gamma^*$ into $\rho^0$) in the ISI.
The degree of shadowing depends on the coherent length $l_c$, which is within the range $0.35\leq l_c\leq 0.8$ fm, as indicated in Table \ref{tb1}.
Since it is less than the case of the $A(e,e'\rho^0)$ reaction, which amounts to $1.35\leq l_c \leq2.45$ fm \cite{aira,dutta107}, such an effect has been neglected in previous studies on this reaction \cite{Qian}.
Nevertheless, it is worth noting that the effect of the shadowing in the initial state is significant at the lower $Q^2$, e.g., 0.55 GeV$^2/c^2$ with the longer $l_c$, while it saturates to a constant at high $Q^2$ \cite{kundu}.

For further understanding of NT extracted from the $A(e,e'\pi^+)$ reaction, we investigate the dependence of $T_A$ on the mass number $A$ in various nuclei with the data taken from Ref. \cite{Qian}.
Figure \ref{fig3} illustrates the $A$ dependence of $T_A$ on a logarithmic scale, which excludes the data point at $A=2$ from the Glauber framework due to the limit of the Wood-Saxon density function applicable to the deuteron target.
A rapid decrease of $T_A$ is shown as the mass number $A$ increases, although the density of nuclei considered here is on average equal regardless of the growing population of nucleons. Thus, the decrease in the $A$ dependence of $T_A$ implies an increase in the in-medium modification, which differs for different nuclei.
From the solid curve with the two-step process on, the shadowing effect can be observed more significantly at the lower $Q^2$, e.g., $0.55$ and 1.1 GeV$^2/c^2$.
The solid curve shows an agreement with data close to the fit of Ref. \cite{Qian} by $1.25A^{\alpha-1}$, which is the best fit of Ref. \cite{Qian} with the least $\chi^2$ value.

The effect of nuclear modification from the free state is studied in Ref. \cite{carroll79}, where the pion-nucleus cross section for the absorption of $\pi^+$ on nuclei is parametrized as $\sigma_A=\sigma_NA^\alpha$ at pion momenta 60, 200, and 280 GeV/c with the parameter $\alpha=0.76$ fitted to empirical data.
When $\alpha=1$, NT is simply a sum in total of $A$ free nucleon contributions , indicating no nuclear medium effect.
Thus, the deviation of the $\alpha$ from unity could be due to the in-medium modification of the mass number $A$.

Meanwhile, in contrast to the $\alpha$ constant in the pion-nucleus scattering \cite{carroll79}, the $\alpha$ obtained from the JLab data for the reaction $A(e,e'\pi^+)$ has a dependence on the photon virtuality $Q^2$, i.e., 0.78 $\leq\alpha(Q^2)\leq$ 0.83, showing the onset of CT due to the role of QDM.
We calculate the $A$ dependence of the transparency $T_A=A^{\alpha(Q^2)-1}$ with the parameter $\alpha$ having $Q^2$ dependence to reproduce JLab data in Fig. \ref{fig4}.
For the global data set of the $A(e,e'\pi^+)$ reaction, we present our prediction by the solid curve, which is comparable with the previous results in Refs. \cite{Larson,Cosyn2} with the $Q^2$ dependence from the QDM up to high $Q^2$, and in particular shadowing also at lower $Q^2$.
As $Q^2$ increases, the departure of $\alpha$ having $Q^2$ dependence from a constant may signify a transition between hadronic and partonic degrees of freedom in nuclear interactions, while quantifying the interplay between color transparency (reducing absorption) and in-medium modifications (enhancing absorption).

\section{Conclusion}

We have studied the NT induced by the subnuclear reaction $p(e,e'\pi^+)n$ from lighter to heavy nuclei, $^{12}$C, $^{27}$Al, $^{63}$Cu, and $^{197}$Au.
The aim of this study is to clarify the reaction mechanism of pion transparency throughout the nuclear medium utilizing the Glauber theory for the nuclear reaction $A(e,e'\pi^+)$ in the range of photon virtuality $0.55\leq Q^2\leq 10$ GeV$^2/c^2$.
Considering QDM in a standard manner, the Glauber theory in the conventional approach could certify CT as an increase in NT proportional to $Q^2$, but it led to an overestimation of the JLab data at 6 GeV electron beams.
To improve the theoretical aspect of the present issue, we introduce the shadowing by a two-step process in the form appropriate for the Glauber theory for the $A(e,e'\pi^+)$ reaction.
We show that the shadowing in the initial state should play an important role in explaining the onset of CT, revealing not only an apparent $Q^2$ dependence that enhances $T_A$ at lower $Q^2\approx 0.5$ GeV$^2/c^2$ but also the $A$ dependence being closer to an agreement with the experimental data.
Now that the two-step process by the intermediate propagation of $\rho^0$ considered in this work is essentially due to the $q\bar{q}$ fluctuation of the virtual photon $\gamma^*$; it is not simply the next-order contribution to the original GT at hadronic level but should be understood as another role of the small-size configuration in the initial state before electroproduction, just as the QDM of the $q\bar{q}$ structure in the final state pion after electroproduction.
We stress that the absorption by shadowing in the ISI in the Glauber theory is indispensable, along with the QDM in the FSI, to provide theoretical guidance for the analysis of electron(lepton)-nuclear reactions in future experiments.

\section*{Acknowledgment}

 This work was supported by the Grant No. NRF-2022R1A2B5B01002307 from the National Research Foundation (NRF) of Korea.

\section*{DATA AVAILABILITY}

 The data supporting this study's findings are available within the article.
 

\end{document}